\documentclass[conference]{IEEEtran}
\IEEEoverridecommandlockouts
\usepackage{cite,hyperref}
\usepackage{amsmath,amssymb,amsfonts}
\usepackage[usenames,svgnames,x11names,dvipsnames]{xcolor}
\usepackage{booktabs,colortbl}
\usepackage{algorithmic}
\usepackage{graphicx}
\usepackage{textcomp}
\usepackage{tabularx}
\usepackage{enumitem}

\begin{document}
\title{Transferable Graph Neural Fingerprint Models for Quick Response to Future Bio-Threats}

\newif\ifblind
\blindtrue
\blindfalse

\author{
\ifblind
    anonymous authors
\else
    \IEEEauthorblockN{%
        Wei Chen\IEEEauthorrefmark{2}\textsuperscript{\textsection},
        Yihui Ren\IEEEauthorrefmark{1}\textsuperscript{\textsection},
        Ai Kagawa\IEEEauthorrefmark{1},
        Matthew R. Carbone\IEEEauthorrefmark{1},
        Samuel Yen-Chi Chen\IEEEauthorrefmark{1},
        Xiaohui Qu\IEEEauthorrefmark{2},
        Shinjae Yoo\IEEEauthorrefmark{1},\\
        Austin Clyde\IEEEauthorrefmark{3},
        Arvind Ramanathan\IEEEauthorrefmark{3},
        Rick L. Stevens\IEEEauthorrefmark{3},
        Hubertus J. J. van Dam\IEEEauthorrefmark{5}, and
        Deyu Lu\IEEEauthorrefmark{2}%
    }%
        \IEEEauthorblockA{\IEEEauthorrefmark{2} \textit{Center for Functional Nanomaterials, Brookhaven National Laboratory, Upton, NY, USA}, dlu@bnl.gov}
    \IEEEauthorblockA{\IEEEauthorrefmark{1} \textit{Computational Science Initiative, Brookhaven National Laboratory, Upton, NY, USA},  yren@bnl.gov}%
       \IEEEauthorblockA{\IEEEauthorrefmark{3} \textit{Data Science and Learning Division, Argonne National Laboratory, Lemont, IL, USA}}
    \IEEEauthorblockA{\IEEEauthorrefmark{5} 
    \textit{Condensed Matter Physics \& Materials Science, Brookhaven National Laboratory, Upton, NY, USA},  hvandam@bnl.gov}%
    \IEEEauthorblockA{\IEEEauthorrefmark{4} \textit{Authors contributed equally.}}
\fi
}


\maketitle

\begin{abstract}
Fast screening of drug molecules based on the ligand binding affinity is an important step in the drug discovery pipeline. Graph neural fingerprint is a promising method for developing molecular docking surrogates with high throughput and great fidelity. In this study, we built a COVID-19 drug docking dataset of about 300,000 drug candidates on 23 coronavirus protein targets. With this dataset, we trained graph neural fingerprint docking models for high-throughput virtual COVID-19 drug screening. The graph neural fingerprint models yield high prediction accuracy on docking scores with the mean squared error lower than $0.21$ kcal/mol for most of the docking targets, showing significant improvement over conventional circular fingerprint methods. To make the neural fingerprints transferable for unknown targets, we also propose a transferable graph neural fingerprint method trained on multiple targets. With comparable accuracy to target-specific graph neural fingerprint models, the training and data efficiency of the transferable model is several times higher. We highlight that the impact of this study extends beyond COVID-19 dataset, as our approach for fast virtual ligand screening can be easily adapted and integrated into a general machine learning-accelerated pipeline to battle future bio-threats.
\end{abstract}


\section{Introduction}
The knowledge of the protein-ligand interaction is essential to many fields in the life sciences, such as biophysics, structural bioinformatics and drug discovery~\cite{sousa2006protein}. Detailed information on the atomic structures and energetics of the protein-ligand complex in the docking conformation is key to unravelling the docking mechanism, which is governed by multiple factors including, e.g., shape matching, electrostatics, hydrogen bonding and van der Waals forces. 

To understand the specific action of a protein on a substrate the lock-and-key model was first proposed by Fischer in 1894~\cite{Fischer1894}. The lock-and-key model proved useful in a variety of contexts including protein-protein interactions as well as protein-ligand interactions. With the emergence of increasingly capable high performance computers, it became possible to automate the search for optimal protein-ligand alignments. This led to the first docking code being developed in 1982~\cite{Kuntz1982}. In addition, it was realized that the lock-and-key model could be used for rational drug design~\cite{Goodford1984}. This realization quickly led to the design and deployment of docking programs specifically for this purpose~\cite{Goodford1985}. Despite remaining challenges, thanks to the growing computing capabilities and improvements in methods and software, docking became a key component in rational drug design.

One outstanding challenge is that the accuracy of the docking results is limited by the approximations required to accelerate the method. 
These approximations involve the way the scoring function accounts for entropy and desolvation effects~\cite{Ferreira2015}. Other limitations stem from the extent to which docking allows for the flexibility of the protein~\cite{Ferreira2015}. These approximations are to a degree due to the scale of the problem that needs to be solved. While docking a single ligand using empirical force field can be done in seconds, the chemical space of drug-like molecules is vast. The size of this space for small molecules with up to 30 atoms has been estimated to be of the order of $10^{60}$ molecules~\cite{Bohacek1996}. While this formidably large chemical space can never be fully explored, smaller but still huge subsets have recently explicitly been considered. The GDB-17 database has 166 billion molecules~\cite{Ruddigkeit2012}; ZINC15 contains over 750 million purchasable compounds~\cite{ZINC15}; ENAMINE enumerates over 22.7 billion compounds, with a database of 4.5 billion {\it REAL} molecules for download~\cite{Enamine-REAL}. Exploring such large molecular spaces is still very expensive, even using fast docking programs, calling for yet more efficient screening techniques. 

One promising path forward is to develop machine learning (ML) models that predict the docking score directly from ligand structures or ligand-target protein structure complexes~\cite{khamis2015machine,crampon2021machine}. Early efforts along these lines have been made since at least 2010 using multiple linear regression, partial least squares regression, random forests, support vector machines, and artificial neural networks~\cite{Ballester2010}. Since then, a variety of machine learning models have been developed, including multiple linear regression~\cite{Ashtawy2015}, multivariate adaptive regression splines~\cite{Ashtawy2015}, gradient boosted trees~\cite{Bucinsky2022,Adeshina2020}, boosted regression trees~\cite{Ashtawy2015}, random forests~\cite{Fernandes2021,Hsin2013}, k-nearest neighbors~\cite{Chandak2020,Ashtawy2015}, support vector machines~\cite{Chandak2020,Ashtawy2015}, logistic regression~\cite{Chandak2020,Hsin2013}, artificial neural networks~\cite{gentile2020deep,Bucinsky2022}, convolutional neural networks~\cite{Ahmed2021,Shen2021deep,McNutt2021}, and graph convolutional neural networks~\cite{Bai2021,Morrone2020}.

Other than the details of the machine learning methods, there are several distinct differences in these models, such as the choice of ground truth in the training set and target applications. For models aimed at generic binding properties, common choices for ground truth on experimental data such as PDBBind~\cite{Liu2014} are used~\cite{Ahmed2021,Ashtawy2015,Hsin2013,Ballester2010}; Binding MOAD~\cite{Hu2005} and Astex~\cite{Hartshorn2007} have also been considered~\cite{Bjerrum2016}. While highly valuable, these experimentally obtained datasets tend to be relatively small. For example PDBBind 2020 contains 19,443 experimentally characterized protein-ligand complexes, which is a small number compared to the possible combinations of more than 188,430 experimental structures in the PDB and billions of theoretically available ligands. An alternative way of obtaining larger ground truth datasets is using computational docking programs to generate simulated datasets. For example, Autodock Vina~\cite{Trott2010,McNutt2021,Morrone2020,Chandak2020} and Gold~\cite{Jones1997,Fernandes2021} have been used to provide ground truth data. The caveat with using these datasets is that they are affected by the same limitations that docking scores in general tend to have. 


Our work has three contributions: 
1) we build a COVID-19 drug docking dataset of about 300,000 drug candidates on 23 coronavirus protein targets;
2) we conduct a systematic study of the popular neural fingerprint models and compare the model performance on this large docking dataset with conventional circular fingerprint models; and
3) we demonstrate the learned neural fingerprints can be used for emerging protein targets under a transfer learning setting.


\section{COVID-19 docking dataset}
Since the first documented case at the end of 2019, coronavirus disease 2019 (COVID-19), caused by the severe acute respiratory syndrome coronavirus 2 (SARS-CoV-2), has quickly evolved into a worldwide pandemic. Unlike previous coronaviruses, such as SARS-CoV and the Middle East Respiratory Syndrome (MERS), SARS-CoV-2 is proved to be significantly more contagious, leading to its exponential spread and a large number of fatalities. With roughly 6.95 million deaths and 768 million infections as of July 2023~\cite{who_covid}, the disastrous impact of this pandemic calls for urgent pharmacological progress in, e.g., vaccines, drugs, and interferon therapies to combat COVID-19~\cite{Ita2021Coronavirus}.

Potential antiviral treatments of SARS-CoV-2 can be divided into two categories acting on either the human immune system or the coronavirus~\cite{Wu2020analysis}. In this work, we focus on the drug molecule docking study of the latter. In general, virus proteins fall into three categories: 1) structural proteins (SPs), which form the virion particles, 2) non-structural proteins (NSPs), which are involved with the virus replication in the host cell, and 3) accessory proteins, which interfere with the host cell's innate immune response~\cite{mariano2020proteins}. A ligand drug molecule may act on SPs to prevent virus from assembling or binding to human cell, or on critical NSPs to inhibit virus RNA synthesis and replication~\cite{Wu2020analysis}.

Since 2020, over 1300 SARS-CoV-2 protein structures (either by themselves or in complexes with other compounds)
have been resolved from experimental facilities around the world.
The atomic structures of these proteins have been determined to high accuracy, which provide the essential structural information of the drug docking sites on SARS-CoV-2. A detailed explanation of the functions of SARS-CoV-2 proteins in the virus life cycle can be found in a recent review~\cite{mariano2020proteins}. Among them, proteins of particular interest are PLPro (an NSP3 domain) and 3CLPro (NSP5), which are the proteases that cut the polyproteins encoded by the viral RNA into active proteins, and ADP Ribose phosphatase (also an NSP3 domain), an innate immune response antagonist. The CoV protein is the receptor binding domain that is the part of the spike protein that binds to the ACE2 receptor triggering the infection. NSP9 is an RNA binding protein that is involved in the viral RNA replication although its precise role is still unclear. NSP10 is the co-factor in the NSP16-NSP10 complex that methylates the cap of newly synthesized RNA, an essential step for RNA stability and function. NSP15 is thought to modify viral RNA at 3' Uracil locations to evade detection by the cell's innate immune system~\cite{Pillon2021}. In addition, researchers leverage the knowledge of the protein functions in other coronaviruses, such as SARS-CoV and MERS as well as the interaction map of SARS-CoV-2 and human proteins \cite{gordon2020sars}. In this work,  we consider 23 pertinent SARS-CoV-2 NSP docking sites as described in Table~\ref{tab:table-targets}.

\begin{table}[thb]
\centering
\caption{\label{tab:table-targets} Description of the 23 SARS-CoV-2 NSP targets. $n_p$ indicates the number of pockets.}
\begin{tabularx}{\columnwidth}{l|c|X}
\toprule
Protein      & $n_p$  & Description \\
\midrule
\hline
3CLPro      & 1  &  3C-like protease\cite{roe2021mpro} \\
ADRP-ADPR   & 2  &  ADP-ribose phosphatase  in complex with ADP ribose\cite{michalska2020adrp}          \\
ADRP        & 3  &  ADP-ribose phosphatase\cite{michalska2020adrp} \\
COV         & 4  &  Receptor binding domain           \\
NSP9        & 2  &  Component of RNA polymerase complex\cite{Kahn2021}   \\
NSP10       & 3  &  Component of 2'-O-RNA methyltransferase complex\cite{Krafcikova2020}  \\
NSP15       & 2  &  Nidoviral RNA uridylate-specific endoribonuclease\cite{Kim2020}           \\
ORF7A       & 1  &   Interferon response antagonist\cite{Cao2021}          \\
PLPro monomer & 3 &  Papain-like protease monomer\cite{rut2020plpro} \\
PLPro dimer      & 2  &  Papain-like protease dimer  \\
\bottomrule
\end{tabularx}
\end{table}

\section{Methods}
\subsection{Docking score prediction workflow}
\begin{figure}[ht]
    \centering
    \includegraphics[width=1.0\hsize]{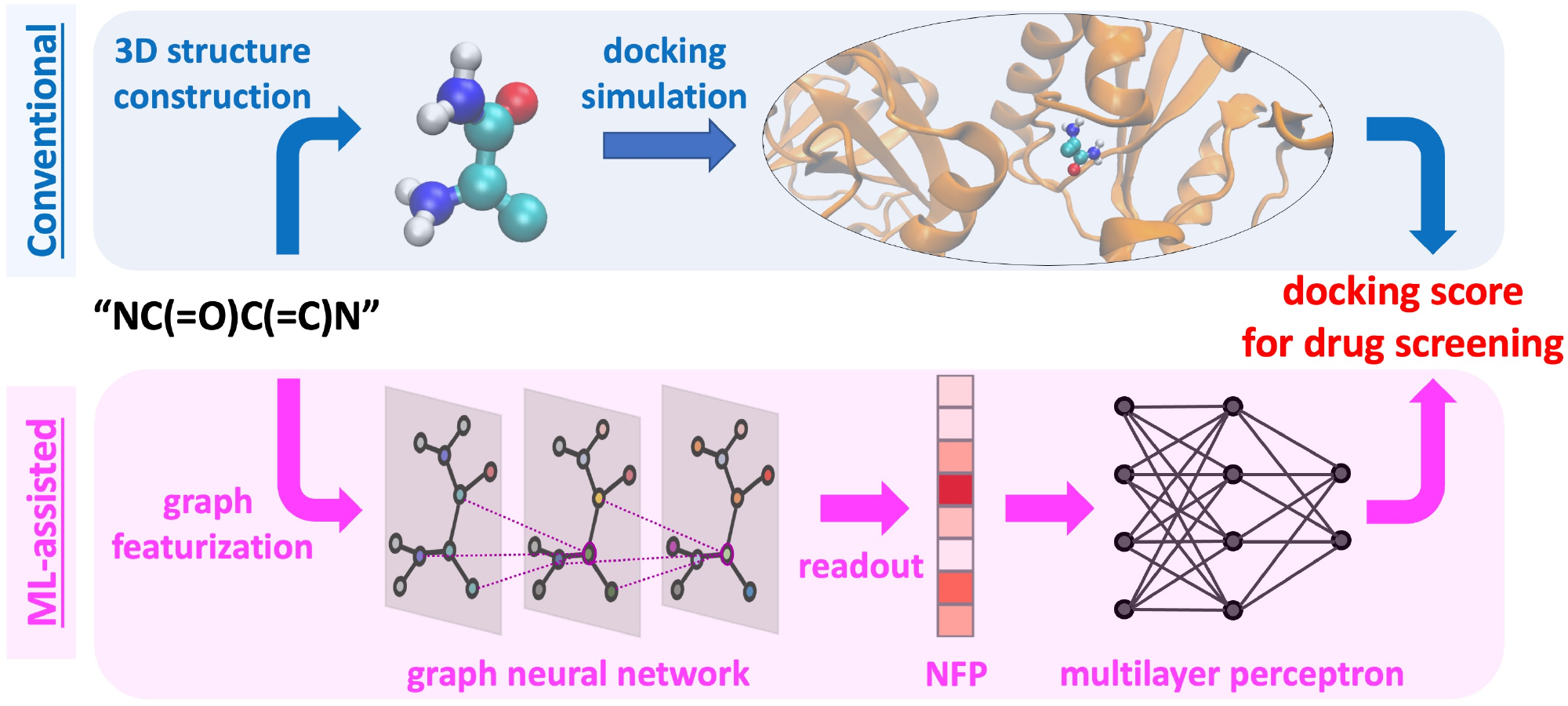}
    \caption{\label{fig:schematic} A schematic flowchart that illustrates two individual processes to compute the docking score of a drug candidate molecule from its SMILES string. The upper route shows the conventional docking simulation, where the docking pose of an exemplary molecule (NC(=O)C(=C)N) on ADRP\_pocket13 is shown. The bottom route shows the graph neural fingerprint model that is trained on the docking simulation data and is able to provide fast and high-fidelity predictions on unknown molecules for drug screening.
    }
\end{figure}

The schematics of our workflow is shown in Fig.~\ref{fig:schematic}, which includes generating a docking dataset from large scale docking simulations (top, shaded in blue) and training surrogate models (bottom, shaded in purple) that can efficiently screen COVID-19 drug candidates in the vast drug-like molecule space. Our ML model predicts the docking scores of drug candidate molecules. While predicting the docking pose using ML models is an exciting open question, it is beyond the scope of this study. The technical details of this workflow are explained below. 

First, we performed docking simulations of 310,693 compounds, including the drug bank compounds, onto each of the 23 coronavirus protein targets using Autodock~\cite{morris2009autodock}. 
The set of targets was generated by using Fpocket~\cite{Guilloux2009} on each protein to identify the top 4 most druggable pockets. The set of pockets was further reduced based on prior experimentally identified binding motifs and visual inspection.
Subsequently all the molecules in the ligand set were processed one-by-one by first converting the molecule's SMILES string to a three dimensional structure using OpenBabel~\cite{OBoyle2011} and the subsequent docking simulations using AutodockTools and Autodock 4.2~\cite{Morris2009}. Autodock 4.2 uses an empirical force field to estimate the ligand binding free energy, which contains pair-wise terms to calculate the interaction between two molecules and an empirical model to estimate contributions from environmental water~\cite{huey2007semiempirical}.
The scoring function in Autodock contains the van der Waals, hydrogen bonding, electrostatic, and desolvation energies~\cite{hill2015scoring}. For each ligand-pocket pair the hybrid genetic algorithm and local search (GA-LS) procedure was executed for 20 times. Each procedure went through maximum 2.5 million energy evaluations to find the lowest energy pose. The final 20 poses were clustered at the end of the AutoDock run and ultimately the best one was selected. The simulation outputs include the optimal docking pose and docking score. The datasets of docking scores are available 
on Github\footnote{See our dataset on \url{https://github.com/BC3D/BC3D_2021}}. 

\vspace{\baselineskip}

\subsection{Machine learning-based surrogate models}
\begin{figure}
    \centering
    \includegraphics[width=0.9 \hsize]{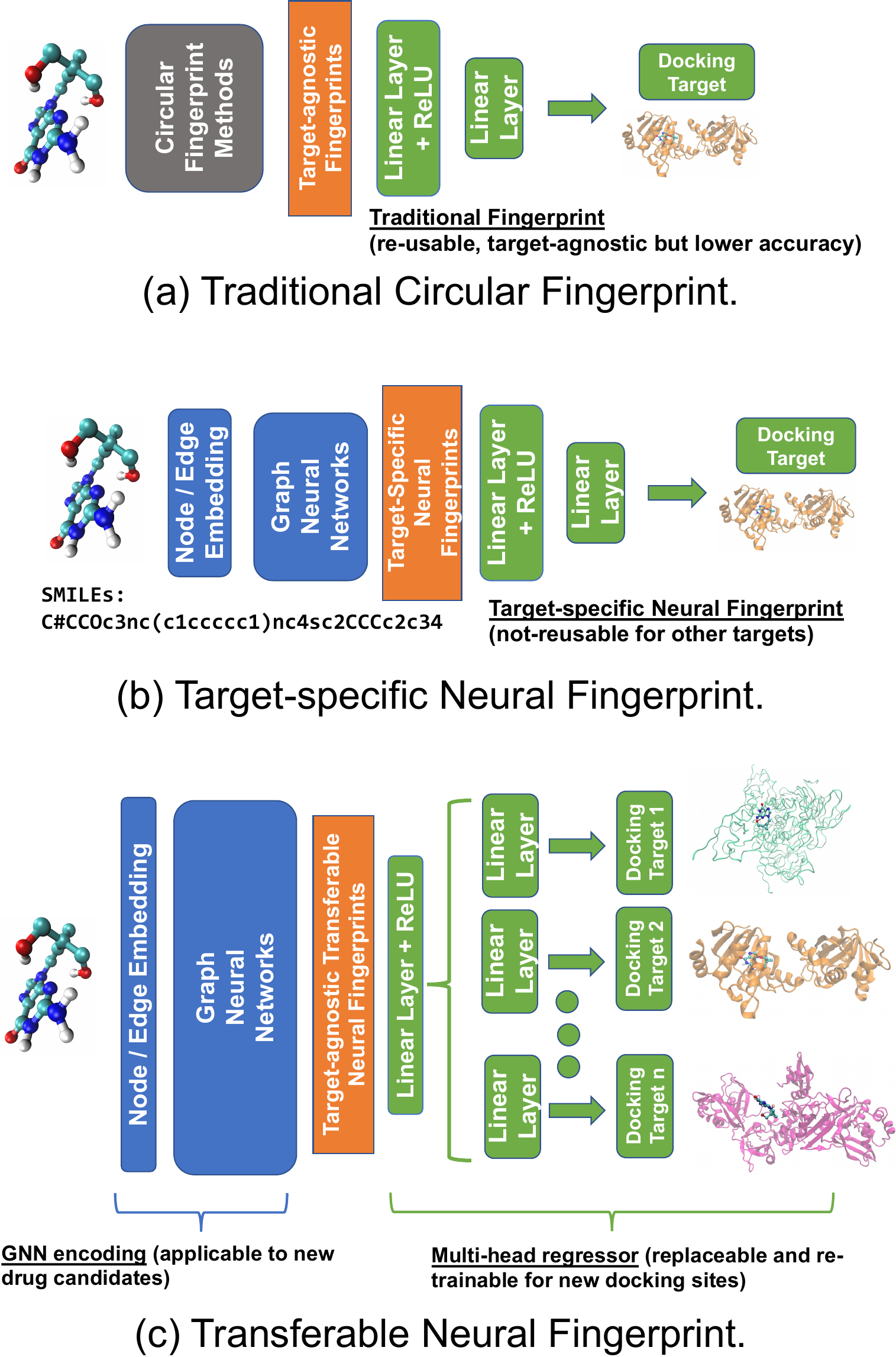}
    \caption{\label{fig:netarch} Comparison of three fingerprinting schemes: traditional circular fingerprint, target-specific neural fingerprint and transferable neural fingerprint.}
\end{figure}

Starting from the SMILES code of molecules, we considered two types of fingerprinting methods based on the featurization of molecules: conventional circular fingerprints (CFP)~\cite{morgan1965generation, rogers2010extended} and neural fingerprints (NFP)~\cite{duvenaud_convolutional_2015-1}. A CFP, such as de Morgan and extended-connectivity fingerprints, abstracts molecular structure information as a vector via hashing, which has been used widely for molecular similarity search for its fast processing time. It is convenient to use CFPs as the input feature of a neural network to perform molecular property predictions, such as the docking score for a particular docking target as shown in Fig.~\ref{fig:netarch}a. Despite the utility for similarity search, the limited structural and chemical information content retained in CFPs impairs their performance for prediction tasks. On the other hand, a NFP-generating model typically has three components: 1) the graph representation and feature embedding, 2) a graph neural network (GNN)-encoder, mapping molecules to a NFP in a fixed-size vector representation, and 3) a multilayer feedforward regressor, e.g., a multilayer perceptron (MLP), to predict a target property. 
These three components are trained together end-to-end as shown in Fig.~\ref{fig:netarch}b. Although it is quite likely that the learned NFP of the same bit-length can outperform the CFP in the docking score prediction task, a systematic comparison is warranted.

In this study, we conducted a systematic study of five types of molecular fingerprinting methods using both CFP and NFP. For conventional CFP methods, we considered de Morgan circular fingerprint (Morgan) and extended-connectivity fingerprint (ECFP) implemented in the open-source cheminformatics software RDKit~\cite{rdkit}. For GNN methods, we considered three popular GNN variations: Gated Graph Convolutional Network (GatedGCN)~\cite{yujia2016gated,bresson_residual_2018}, GraphSAGE~\cite{ying_hierarchical_2018,hamilton_inductive_2017} and a boilerplate Message-Passing Neural Network (MPNN)~\cite{gilmer_neural_2017}. We ensured the fingerprints have the same bit-length of 2048. Specifically, the CFPs have a fixed length of 2048 bits, while the NFPs are represented by a vector of 16-bit floating points of length 128. We found that the NFP model performance remained the same when we reduced the NFP length from 128 to 70 in GatedGCN and GraphSAGE models. Then both types of fingerprints were fed into a 3-layer perceptron for target-specific regression as shown in Figs.~\ref{fig:netarch}a and~\ref{fig:netarch}b, except for the MPNN where we used only a single layer perceptron.



To process molecular structures into features suitable for GNN training, we first converted the molecular SMILES~\cite{weininger_smiles_1988} representation to a multi-attribute 
undirected graph representation, where nodes and edges correspond to atoms and chemical bonds, respectively, as shown in Figs.~\ref{fig:schematic} and \ref{fig:netarch}b. We followed Gilmer \textit{et al.}~\cite{gilmer_neural_2017} to encode the chemical attributes, such as atomic types and bond types, in node and edge features, unless otherwise stated.

Such categorical features were then mapped to short trainable vectors of real values known as feature embeddings. Therefore, each molecule can be represented as a triplet of matrices of $(\mathbf{A}, \mathbf{V}, \mathbf{E})$, where $\mathbf{A}$ is the adjacency matrix of a graph with added
self-loops, and $\mathbf{V}$ and $\mathbf{E}$ are node and edge embedding matrices, respectively.

We explored various GNNs~\cite{duvenaud_convolutional_2015-1,kipf_semi-supervised_2016} to encode a graph topology $\mathbf{A}$, its node embedding $\mathbf{V}$ and edge embedding $\mathbf{E}$ into a fixed-length fingerprint. GNNs are among the best deep learning models for handling networked data due to its permutation invariance property. Namely, the order of nodes represented in an adjacent matrix does not affect the prediction. Most GNN models consist of an iterative sequence of alternating two steps: \textit{communication} and \textit{aggregation}. During communication, each node will gather its neighboring node embeddings and incident edge embeddings of previous step. During aggregation, trainable functions (e.g. neural networks) are applied to these embeddings, aggregated into a single vector and used to update the node's existing embedding. The same procedure applies to the edge embedding updates.  Different types of GNNs differ in the functions and procedures applied to the neighboring embeddings and the type of aggregation used.

\begin{itemize}[leftmargin=*]
\item Our MPNN model is a slight modification of that presented in Gilmer \textit{et al.}~\cite{gilmer_neural_2017}. It is implemented using Deep Graph Library~\cite{wang_deep_2019}, where atoms are featurized according to the Weave atom featurizer~\cite{kearnes2016molecular}. Messages are learned by a MLP, which are passed between neighboring atoms and used to update both the node and edge embeddings using a Gated Recurrent Unit.

\item GraphSAGE~\cite{yujia2016gated,bresson_residual_2018} is a stochastic generalization of graph convolutions which features sampling and hierarchical aggregation. The aggregated embedding is concatenated with that of the central node before a fully-connected layer. We used a max pooling approach as the aggregation function.

\item In the Gated-GCN network~\cite{yujia2016gated,bresson_residual_2018}, before aggregation each neighboring node embedding is gated (softmax) by a trainable linear combination of both nodes. 

\end{itemize}

\subsection{Transferable neural fingerprint}
CFP and NFP have an important distinction in the nature of their fingerprints (see Figs.~\ref{fig:netarch}a and \ref{fig:netarch}b). The molecular fingerprint in CFP is derived solely from the molecular structure, making it target agnostic and reusable. In contrast, the fingerprint used in the single target NFP model (Fig.~\ref{fig:netarch}b) is target specific, as they are trained as part of the neural network with the knowledge of the docking scores of a given target. As a result, the task-specific NFP approach requires re-training for different docking targets. Therefore, \textit{standard NFP methods lack the most prominent advantage of the CFP: pre-computable and target-agnostic}. This drawback severely limits NFP's practical  utility, as for each new protein target it has to be retrained on large amount of docking data and the fingerprint database grows with the number of protein targets, comparing to the case of CFP where the database is invariant to new protein targets. On the physical ground, the docking simulation searches for the lowest energy configuration of the ligand under the given potential field (e.g., electrostatics, van der Waals and hydrogen bond) created by the binding pocket. Therefore, in principle a molecular fingerprint is transferable, as far as it is featurized to encode essential chemical attributes (e.g., atomic charge, van der Waal radius and interaction strength, and the location of hydrogen-bond donor or acceptor) of the ligand that determine the ligand-pocket interaction energy under the target's potential field. 

To this end, we propose transferable neural fingerprints (TNFPs) that combine the benefits of both CFPs and NFPs. On one hand, TNFPs are re-computable and target-agnostic like CFPs; on the other hand, they encode more complete molecular structural information like NFPs. As shown in Fig.~\ref{fig:netarch}c, we trained TNFPs via a multi-target model. The learned TNFP can be stored in a database, and the GNN encoder can extract a TNFP from newly added drug molecules. For the newly identified docking target, we can train a dedicated MLP regressor  starting from TNFP, which is a much faster and more data-efficient (i.e., requiring less training data) process. To make a fair comparison, all the fingerprints in this study have the same bit length of 2048, either 2048 bits as in CFP or 128 float-16 in NFPs and TNFPs.

\section{Analysis of the Drug-like Ligand Docking Score Dataset}
Our raw dataset contains molecular docking results (atomic coordinates for each ligand-target pair from its lowest-energy docked conformation and the corresponding docking score) of 310,693 ligands on each of 23 targets.  First we conducted a data cleaning process to retain data relevant to drug screening. Out of the 23 targets, 5 of them (i.e., NSP10\_pocket1, NSP10\_pocket3, NSP10\_pocket26, NSP15\_pocket2, and PLPro\_chainA\_pocket4) show positive docking scores on nearly all ligands (more than 99.8\%). These target were removed from the dataset, due to the overall low drug affinity. We also removed about 1K non-drug ligands that are either too large (e.g., protein and RNA) or too small (e.g., salt), as well as nearly 5K non-bonded ligands that contain disconnected parts. After the cleaning process the final dataset includes docking scores of 300,457 ligands on each of 18 different druggable targets.

\begin{figure}[htb]
    \centering
    \includegraphics[width=0.9\hsize]{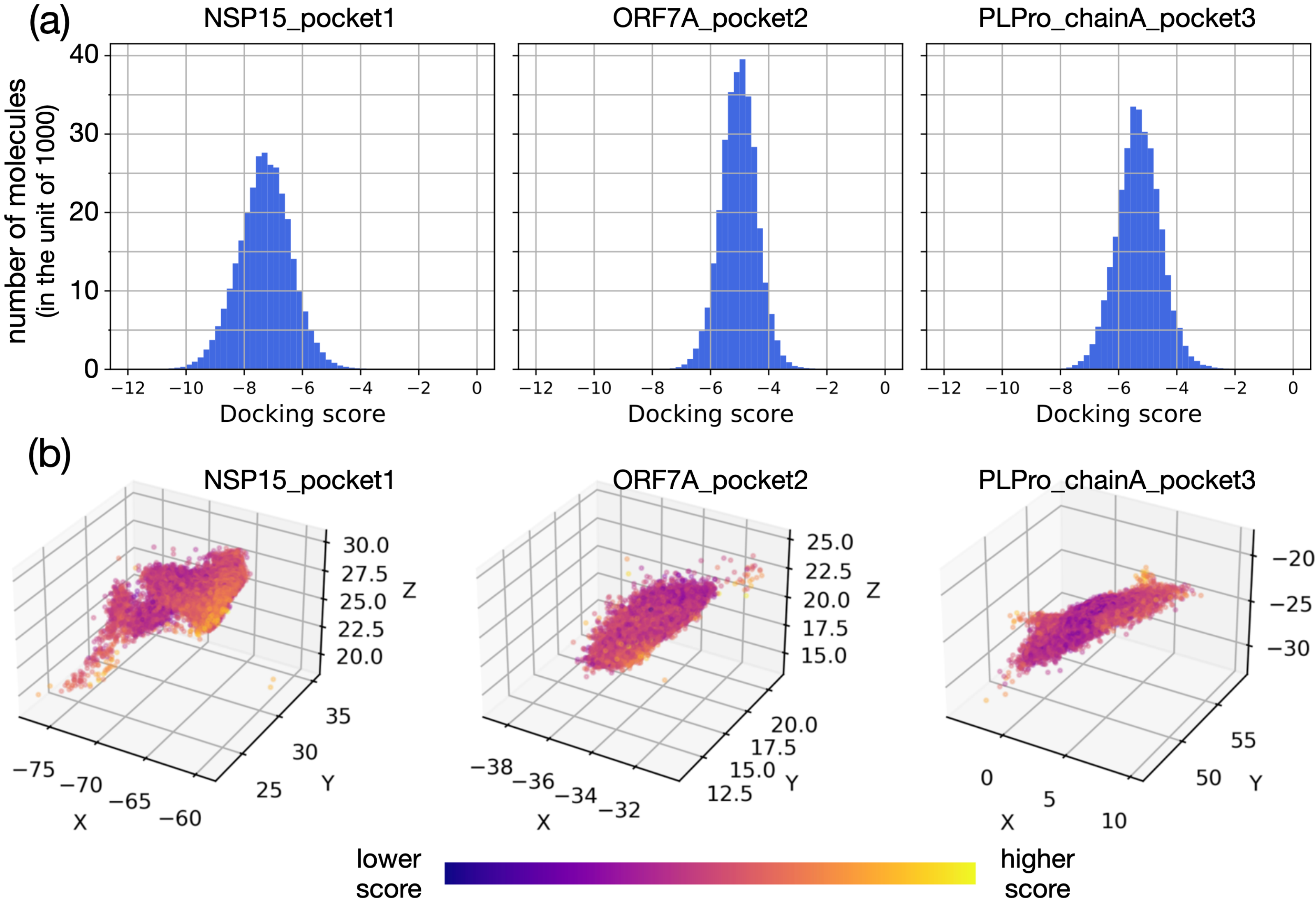}
    \caption{\label{fig:data} (a) Docking score distributions and (b) center-of-mass coordinate distributions of molecules for three representative docking targets. The docking scores are in units of kcal/mol and the coordinates in angstrom.
    }
\end{figure}

Next we examined the data variance from two perspectives. First, on each target the ligand molecules exhibit a broad range of docking scores (Fig.~\ref{fig:data}a and the supporting information Fig.~S1). The docking score distribution is typically Gaussian-like with a single peak, except for one target -- 3CLPro\_pocket1, which shows a bimodal distribution. Second, across different targets the peak position and width of the distribution vary quite significantly. The latter can be also seen from the variation of the average score (i.e., averaged over either all molecules or top 100 molecules in the lower end of the docking score) across the 18 targets (see the supporting information Fig.~S2). Since a more negative docking score means stronger ligand-target affinity, targets with more negative tails, including ADRP-ADPR\_pocket5, ADRP-ADPR\_pocket1, and PLPro\_pocket50, are likely promising druggable sites.

To understand the bimodal distribution in 3CLPro\_pocket1, we further examined the docking configurations of the ligand-target pairs. Given more than 300K molecules in the docking simulations, we computed the center-of-mass (COM) distribution of molecules as a proxy of their docking configurations (see Fig.~\ref{fig:data}b and the supporting information Fig.~S3). 3CLPro\_pocket1 is clearly an exception, because its ligand COM distribution shows disconnected regions in space. We sampled and examined structures from each region and found that the pre-defined docking area in 3CLPro\_pocket1 is highly solvent-exposed and consisted of multiple dockable regions. Such observations indicated that the docking area in 3CLPro\_pocket1 may be ill-defined. In contrast, the spatial distributions in all other targets are continuous despite of irregular shape in several targets. 

\begin{figure}[htb]
    \centering
    \includegraphics[width=\columnwidth]{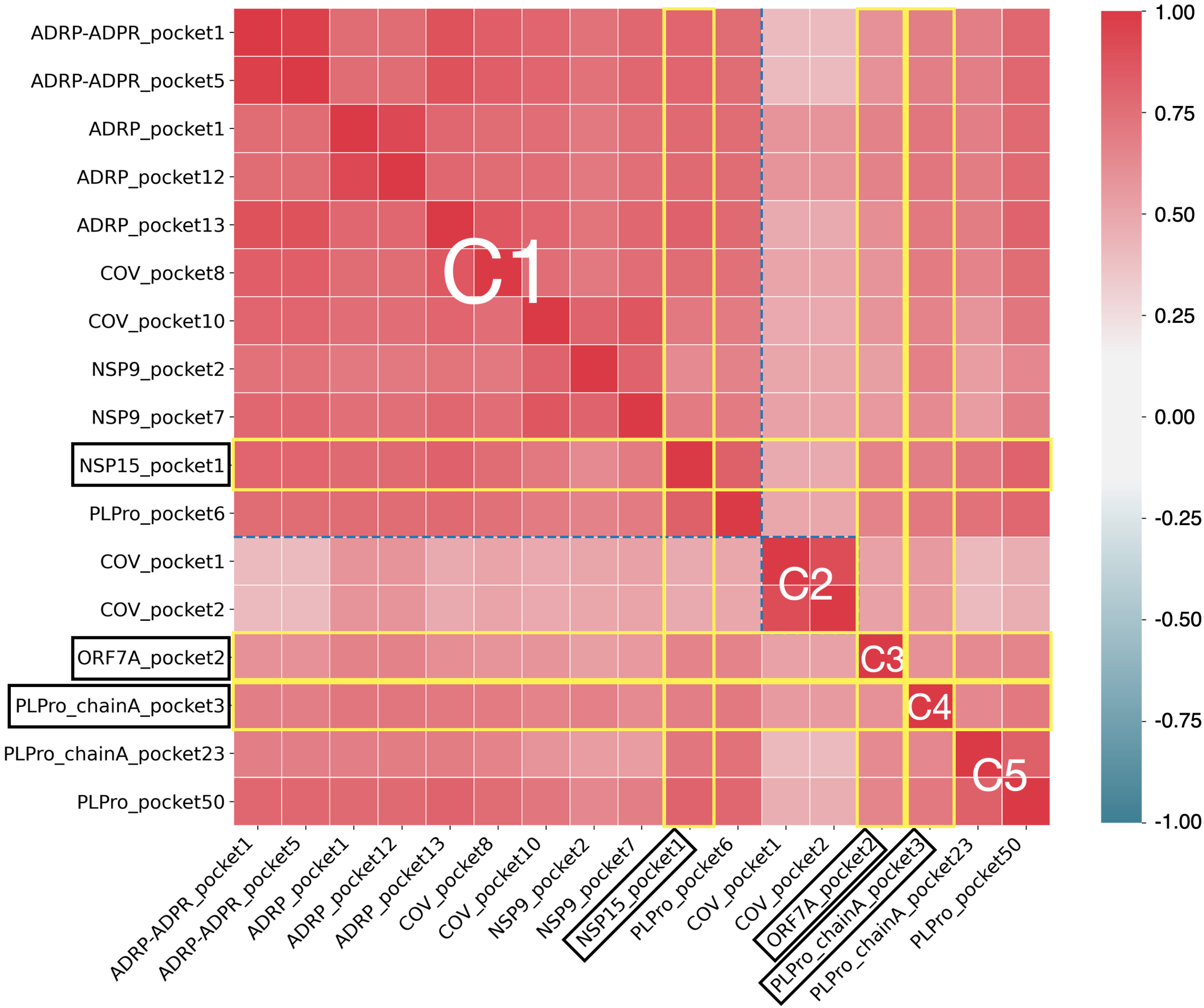}
    \caption{\label{fig:corr} Hierarchical clustering of targets based on the correlation matrix that measures the similarities between targets. The color indicates the value of the Pearson's correlation coefficient. Five clusters are labeled as C1-C5, indicated by blocks separated by dashed lines.  Targets within the same block belong to the same cluster. Boxed target names indicate the targets chosen to test our multi-target model.
    }
\end{figure}

To measure the similarities among targets, we first defined a 300457-dimension vector for each target that contains docking scores of all the ligand molecules. Then we calculated the Pearson's correlation coefficients and performed a hierarchical clustering analysis on the correlation matrix using the scikit-learn package~\cite{sklearn_api} (see Fig.~\ref{fig:corr}). The targets are all positively correlated, namely, a molecule binds strongly (weakly) to one target is also more likely to bind to other targets strongly (weakly). Based on the correlation matrix the targets were grouped into five clusters (C1 to C5).  The largest cluster in the upper left of Fig.~\ref{fig:corr} contains 11 targets, while the rest of the clusters have at most 2 targets. Within each cluster the targets share significant similarities.


\section{Results and Discussions}
We trained two types of target-specific GNN models to predict docking scores of drug candidate molecules  -- one based on CFP (Fig.~\ref{fig:netarch}a) and the other based on NFP of molecules (Fig.~\ref{fig:netarch}b). The performance of the prediction on the test set was reported in Table~\ref{tab:table-performation}. As a reference for comparison, the mean squared error (MSE) of the baseline model was also reported, which refers to the error with respect to the average score in the training set. More information about the architectures and configurations of our graph neural networks is explained in the supporting information.

First, all GNN models show significant improvement over the baseline by 2- to 10-fold, suggesting that GNN-based surrogate models can indeed capture the non-trivial correlation between the molecular structure of the ligand and its docking score on a specific target. Second, NFP-based models outperform the CFP-based models systematically with smaller MSE values by 0.01 $\sim$ 0.08. Third, the three NFP-based models perform equally well with minor MSE differences (smaller than 8\%) between them, and the GatedGCN model is the best in most cases. We included a scatter plot of the ground truth versus prediction using the GatedGCN model in the supporting information Fig.~S4. We noticed that the prediction of docking scores of molecules on 3CLPro\_pocket1 is the worst with the MSE an order of magnitude higher than the rest. This large error is likely caused by the fact that 3CLPro\_pocket1 does not have a well-defined docking pocket as discussed in the above section.

When models are used to perform drug screening, the top ranked (e.g., top 10\%) molecules are usually selected for further testing. In this regard the rank correlation between molecules can be more important than the error on an individual molecule when a model is evaluated. A commonly used metric to measure the rank correlation is the concordance index (CI)~\cite{ozturk2018deepdta}. In our GatedGCN models, the CIs are between 0.76 and 0.91 (see Fig.~S4). 

\begin{table*}[htb]
\begin{center}
\caption{\label{tab:table-performation} MSEs of the docking score prediction on the test sets from two CFP models (ECFP and Morgan) and three graph-based models (GatedGCN, GraphSAGE, and MPNN). The best performer for each target is indicated in bold font. The baseline model corresponds to the naive variations based on the average docking score in the training sets.}
\begin{tabular}{l|r|rr|rrr}
\toprule
                Target &  Baseline &   ECFP &  Morgan &  GatedGCN &  GraphSAGE &   MPNN \\
\midrule
\hline
        3CLPro\_pocket1 &     1.830 &  1.132 &   1.115 &\bf{1.031} &      1.096 &  1.061 \\
     ADRP-ADPR\_pocket1 &     1.588 &  0.237 &   0.202 &\bf{0.151} &      0.157 &\bf{0.151}\\
     ADRP-ADPR\_pocket5 &     1.584 &  0.237 &   0.200 &\bf{0.149} &      0.156 &  0.153 \\
          ADRP\_pocket1 &     0.544 &  0.115 &   0.109 &\bf{0.085} &      0.092 &  0.087 \\
         ADRP\_pocket12 &     0.544 &  0.116 &   0.109 &\bf{0.085} &      0.091 &  0.086 \\
         ADRP\_pocket13 &     1.048 &  0.163 &   0.144 &\bf{0.104} &      0.111 &\bf{0.104} \\
           COV\_pocket1 &     0.270 &  0.076 &   0.070 &\bf{0.054} &      0.058 &  0.058 \\
           COV\_pocket2 &     0.271 &  0.075 &   0.069 &\bf{0.055} &      0.058 &  0.057 \\
           COV\_pocket8 &     0.872 &  0.178 &   0.162 &\bf{0.125} &      0.133 &  0.127 \\
          COV\_pocket10 &     1.166 &  0.172 &   0.163 &\bf{0.120} &      0.124 &  0.124 \\
          NSP9\_pocket2 &     1.139 &  0.205 &   0.211 &\bf{0.164} &      0.170 &  0.167 \\
          NSP9\_pocket7 &     0.987 &  0.135 &   0.126 &\bf{0.089} &      0.093 &  0.092 \\
         NSP15\_pocket1 &     0.843 &  0.169 &   0.156 &\bf{0.122} &      0.126 &\bf{0.122} \\
         ORF7A\_pocket2 &     0.397 &  0.149 &   0.144 &\bf{0.123} &      0.127 &  0.124 \\
  PLPro\_chainA\_pocket3 &     0.566 &  0.167 &   0.161 &\bf{0.134} &      0.142 &  0.135 \\
 PLPro\_chainA\_pocket23 &     0.927 &  0.253 &   0.242 &     0.199 &      0.207 &\bf{0.195} \\
         PLPro\_pocket6 &     0.843 &  0.157 &   0.139 &     0.112 &      0.118 &\bf{0.110} \\
        PLPro\_pocket50 &     1.335 &  0.293 &   0.272 &\bf{0.211} &      0.222 &  0.215 \\
\bottomrule
\end{tabular}
\end{center}
\end{table*}
Next we trained a multi-target model via a GNN encoder with multi-head regressors, one for each docking site, as shown in Fig.~\ref{fig:netarch}c. We used 14 targets for training and tested the transferability of the learned TNFP on the remaining targets. We excluded 3CLPro\_pocket1 for this task based on the docking score and configuration analysis in the previous section. Based on the similarities among targets (Fig.~\ref{fig:corr}), we chose NSP15\_pocket1, ORF7A\_pocket2, and PLPro\_chainA\_pocket3 as the test targets such that they belong to different target clusters. The first target was randomly picked, and the latter two were intentionally chosen to minimize the similarities to the 14 training targets. For the purpose of comparison, we included the results of the single-target GatedGCN and CFP models that were adopted from Table~\ref{tab:table-performation}. On the three test targets, the TNFP model performance shows a noticeable improvement over the CFP models, and is slightly worse than that of GatedGCN (difference within 11.5\%), as shown in the supplementary information Fig.~S5.

The transferable nature of the TNFP model gives it the advantages in training efficiency and data efficiency. The training efficiency is reflected in the runtime of each model. On average the TNFP model costs 53 seconds/epoch on our GPU node while the GatedGCN single-target model costs almost twice (90 seconds/epoch) and the CFP models about 9 times (410 seconds/epoch) as much as the TNFP model. The training effiency of TNFP arises from two factors. First, its input dimension, i.e., the number of nodes in the input layer, is much smaller than the CFP models. Second, its graph encoding part is fixed (i.e., transferable) and does not require re-training as opposed to the single-target NFP model.

\begin{figure}[th]
    \centering
    \includegraphics[width=0.9\hsize]{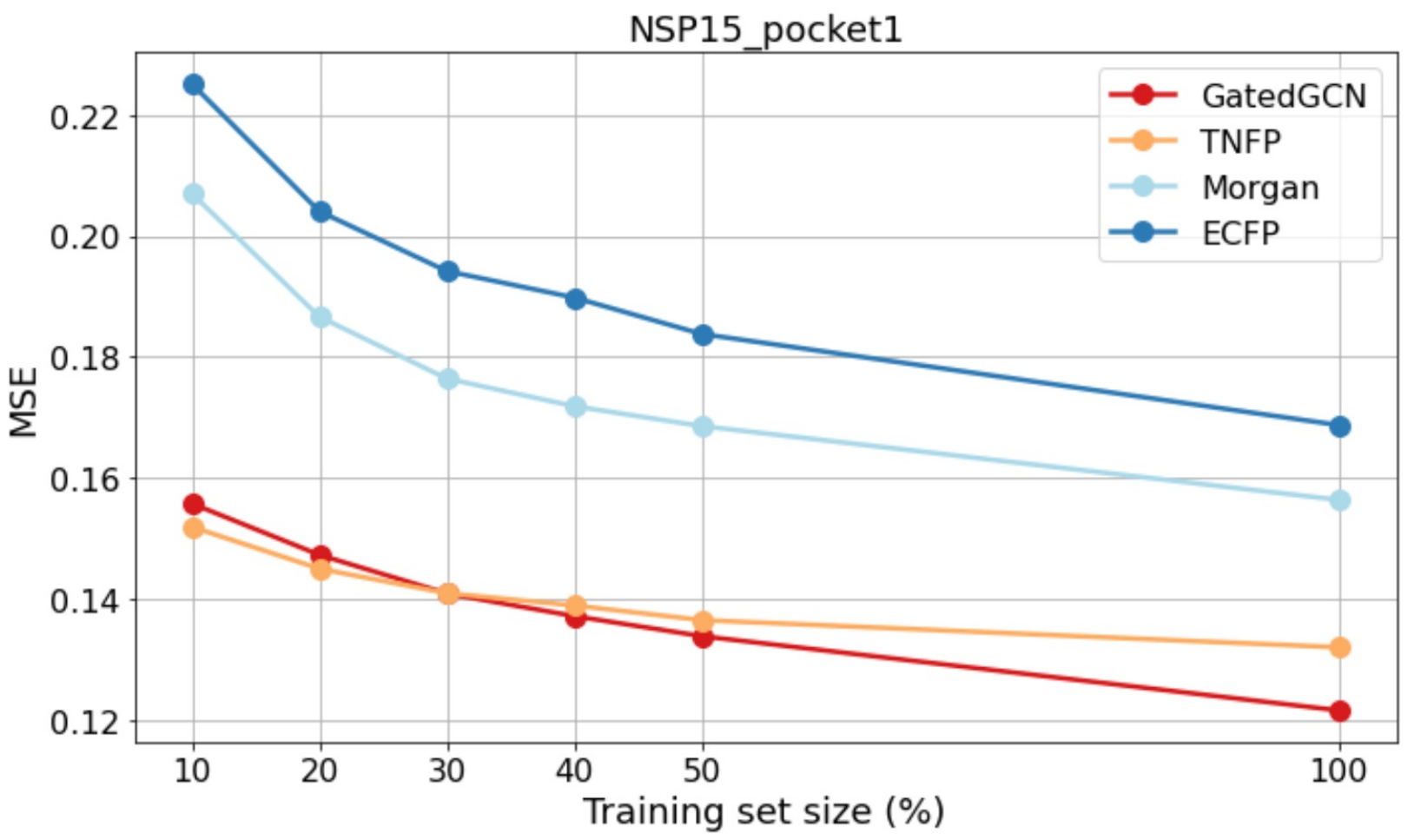}
    \caption{\label{fig:data_eff} Training data efficiency of four different models on NSP15\_pocket1.}
\end{figure}

The data efficiency can be extremely useful when the training data is limited, which is often the case for newly discovered protein targets. As we reduced the training size from the original size (240,000 molecules), the MSE on the test set (size remained 30,457) increases much faster in the GatedGCN model and the two CFP models than the TNFP model (see Fig.~\ref{fig:data_eff} for NSP15\_pocket1 and Fig.~S6 for the other two test targets). As we see in Fig.~\ref{fig:data_eff}, below the crossover at 30\%, the TNFP model outperforms the GatedGCN model. With 10\% of the training data (24,000 molecules), the MSE of the TNFP model increases by only 15.1\% as compared to that of the full tranining set, while the MSE of the GatedGCN model increases dramatically by 28.1\%.

\section{Conclusions}
In summary, we have built and analyzed a COVID-19 docking datasets consisting of $\sim 3\times 10^5$ drug candidates and 23 coronavirus protein targets using Autodock.
We have conducted a comprehensive study of various graph neural network methods to construct surrogate models for  docking score prediction, including both conventional circular fingerprint methods (ECFP and Morgan) and graph neural fingerprint methods (GatedGCN, GraphSAGE, and MPNN). 
We found that overall graph neural fingerprint methods outperform the conventional circular fingerprint methods with the same bit-length of 2048, and GatedGCN performs slightly better than GraphSAGE and MPNN. 
However, graph neural fingerprint methods are target specific and require re-training for new docking targets, which makes them more data intensive and computationally more expensive to train than the conventional circular fingerprint methods. By withholding five representative protein targets as unknown emerging bio-threat , we demonstrated that the  neural fingerprints learned via multi-target training exhibits desired target agnostic and reusable properties of circular fingerprints.
We found that the transferable graph neural fingerprint model not only outperforms  conventional circular fingerprint models, but also shows outstanding training and data efficiency.

\section{Acknowledgments}
This research was supported by the DOE Office of Science through the National Virtual Biotechnology Laboratory, a consortium of DOE national laboratories focused on response to COVID-19, with funding provided by the Coronavirus CARES Act and as part of the CANDLE project by the DOE-Exascale Computing Project (17-SC-20-SC).
This research used the theory and computation resources of the Center for Functional
Nanomaterials, which is a U.S. DOE Office of Science Facility,
and the Scientific Data and Computing Center, a component
of the Computational Science Initiative, at Brookhaven
National Laboratory under contract no. DE-SC0012704.
M.R.C. acknowledges the support by the U.S. Department of Energy, Office of Science, Office of Advanced Scientific Computing Research, Department of Energy Computational Science Graduate Fellowship under Award Number DE-FG02-97ER25308.
This research used resources of the Argonne Leadership Computing Facility, which is a DOE Office of Science User Facility supported under contract DE-AC02-06CH11357.
W.C. would like to thank Yue Qian and Mark S. Hybertsen for helpful discussions.


\bibliographystyle{IEEEtran}

\bibliography{gnn}

\end{document}


\title{Supplemental Material for Transferable Graph Neural Fingerprint Models for Quick Response to Future Bio-Threats}


\author{Wei Chen}
\thanks{Contributed equally to this work}
\affiliation{Center for Functional Nanomaterials, Brookhaven National Laboratory, Upton, New York 11973}

\author{Yihui Ren}
\thanks{Contributed equally to this work}
\affiliation{Computational Science Initiative, Brookhaven National Laboratory, Upton, New York 11973}

\author{Ai Kagawa}
\affiliation{Computational Science Initiative, Brookhaven National Laboratory, Upton, New York 11973}

\author{Matthew R. Carbone}
\affiliation{Computational Science Initiative, Brookhaven National Laboratory, Upton, New York 11973}

\author{Samuel Yen-Chi Chen}
\affiliation{Computational Science Initiative, Brookhaven National Laboratory, Upton, New York 11973}

\author{Xiaohui Qu}
\affiliation{Center for Functional Nanomaterials, Brookhaven National Laboratory, Upton, New York 11973}

\author{Shinjae Yoo}
\affiliation{Computational Science Initiative, Brookhaven National Laboratory, Upton, New York 11973}

\author{Austin Clyde}
\affiliation{Computational Science \& Data Science and Learning Division, Argonne National Laboratory, Lemont, Illinois 60439}

\author{Arvind Ramanathan}
\affiliation{Data Science and Learning Division, Argonne National Laboratory, Lemont, Illinois 60439}

\author{Rick L. Stevens}
\affiliation{Computing, Environment and Life Sciences, Argonne National Laboratory, Lemont, Illinois 60439}

\author{Hubertus J. J. van Dam}
\email{hvandam@bnl.gov}
\affiliation{Condensed Matter Physics and Materials Science, Brookhaven National Laboratory, Upton, New York 11973}

\author{Deyu Lu}
\email{dlu@bnl.gov}
\affiliation{Center for Functional Nanomaterials, Brookhaven National Laboratory, Upton, New York 11973}

\maketitle
\appendix

\setcounter{figure}{0}
\setcounter{table}{0}

\makeatletter 
\renewcommand{\thefigure}{S\@arabic\c@figure}
\renewcommand{\thetable}{S\@arabic\c@table}  
\makeatother


\section{GCN and GraphSAGE Model architectures}

\noindent
The node and edge embedding procedures map 
from vectors in the feature dimensions
to vectors in the embedding dimensions respectively for nodes and edges.  The node and edge feature and embedding dimensions are shown in the following tables.

\vspace{.3cm}

\begin{tabular}{|l|l|}
\hline
Node feature dimensions   & {[}11, 7, 9, 8, 6, 3{]}     \\ \hline
Node embedding dimensions & {[}15, 15, 15, 10, 10, 5{]} \\ \hline
\end{tabular}

\vspace{.3cm}

\begin{tabular}{|l|l|}
\hline
Edge feature dimensions   & {[}5, 5{]}   \\ \hline
Edge embedding dimensions & {[}35, 35{]} \\ \hline
\end{tabular}

\vspace{.3cm}
\noindent
The output of the universal neural fingerprint layers is the input of the fully connected, multilayer perceptron (MLP). 
This MLP consists of two layers, and the number of nodes in each layer
and activation functions are listed in the following tables.
ReLU is Rectified Linear Unit.

\vspace{.3cm}
\begin{tabular}{ | p{4.5cm} | p{3cm} | }
\hline
Number of nodes from inputs to outputs   
& {[} 70, 35, 1{]}  \\ \hline
Activation functions from the first layer to last layer 
& {[} ReLU, None {]}  \\ \hline
\end{tabular}

\section{Details on training the machine learning models}
The docking dataset was randomly split into 240,000, 30,000, and 30,457 samples for training, validation, and testing, respectively, using the \emph{numpy.random.shuffle} function~\cite{harris2020array}. The validation set was used to optimize hyperparameters including the numbers of hidden layers and neurons as well as perform early stopping (maximum 300 epochs) and model selection. The mean squared error (MSE) was used as the loss function. The models were trained using the ADAM optimizer~\cite{kingma2014adam} with batch size 64. The initial learning rate was 0.001 with a reducing factor of 0.7 and a minimum learning rate of $10^{-5}$. This ML experiments were conducted using one node of the BNL institutional cluster. Our code used 20 Intel Xeon Gold 6248 processors and a NVIDIA V100 GPU on one node of the cluster.  Each experiment for a particular target takes about 5-6 hours.

\begin{figure}
    \centering
    \includegraphics[width=0.9\hsize]{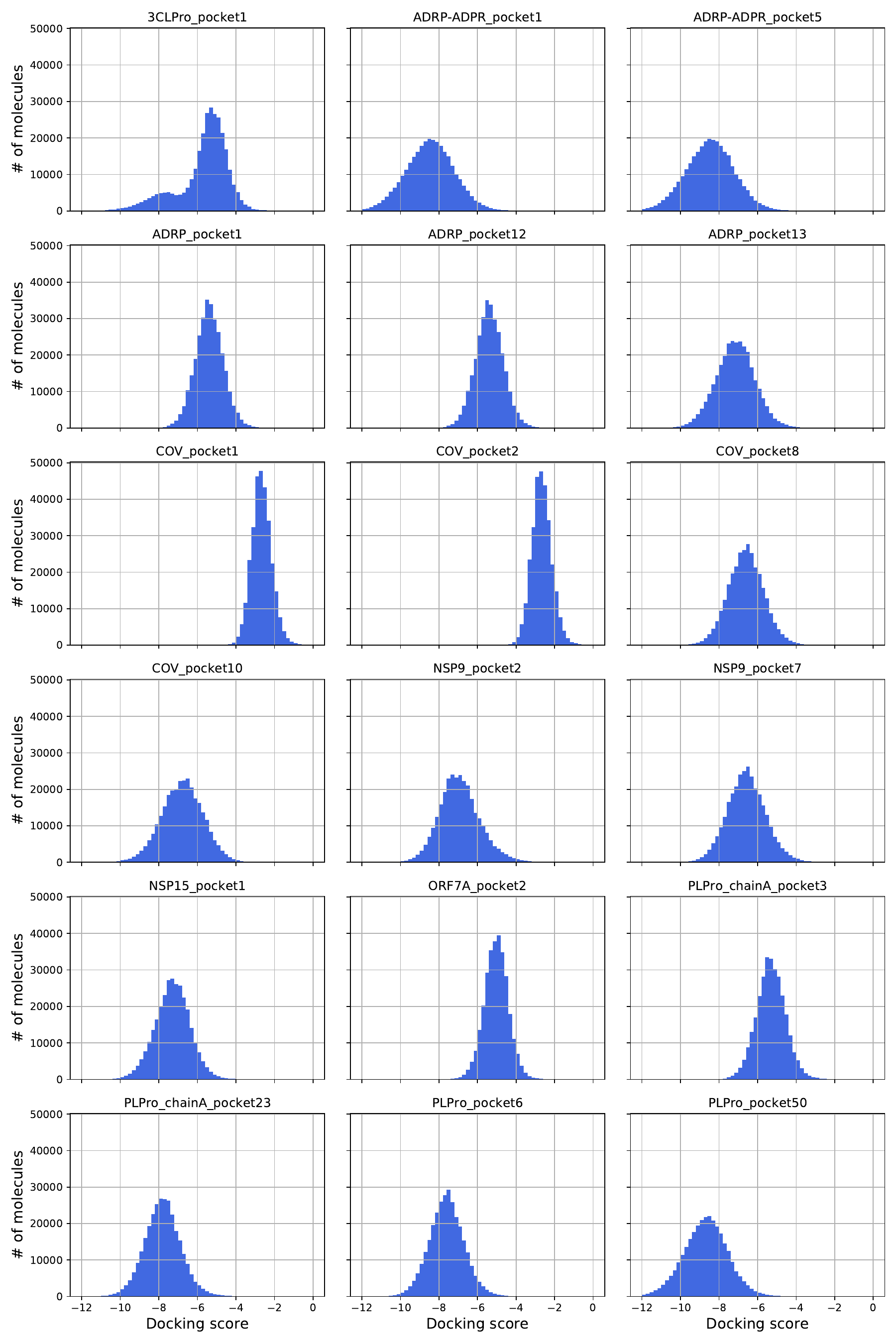}
    \caption{\label{fig_s:dist} Docking score distributions of molecules for 18 docking targets. The scores are in units of kcal/mol.
    }
\end{figure}

\begin{figure}
    \centering
    \includegraphics[width=\hsize]{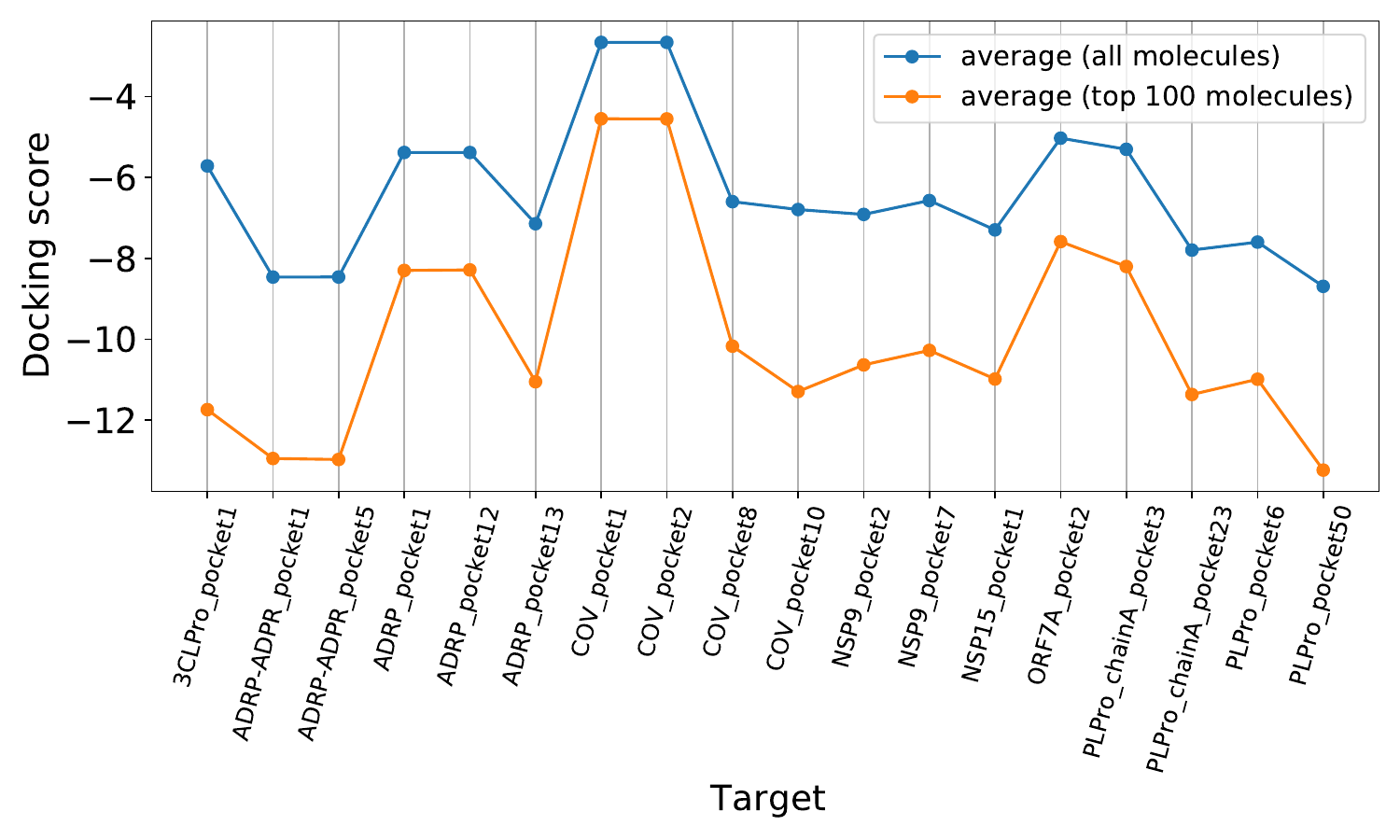}
    \caption{\label{fig_s:avg} Average docking scores of molecules for 18 docking targets. 
    }
\end{figure}

\begin{figure}[b!]
    \centering
    \includegraphics[width=\hsize]{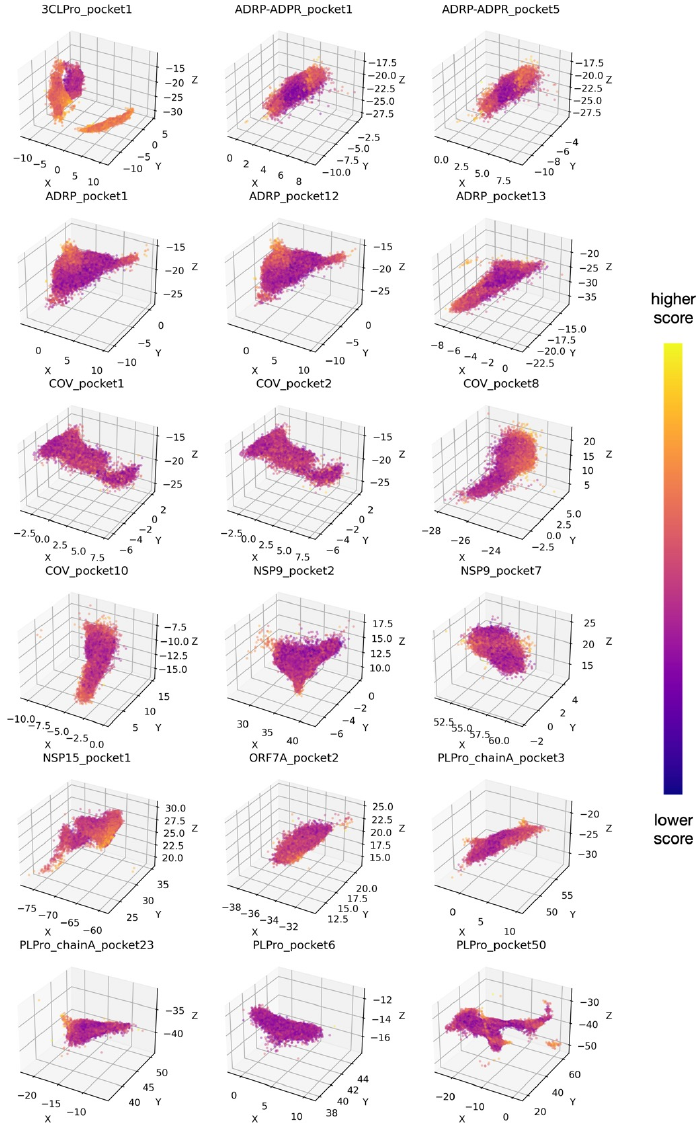}
    \caption{\label{fig_s:com} Center-of-mass distributions of molecules for 18 docking targets. The coordinates are in units of angstrom.
    }
\end{figure}

\begin{figure}[t!]
    \centering
    \includegraphics[width=0.9\hsize]{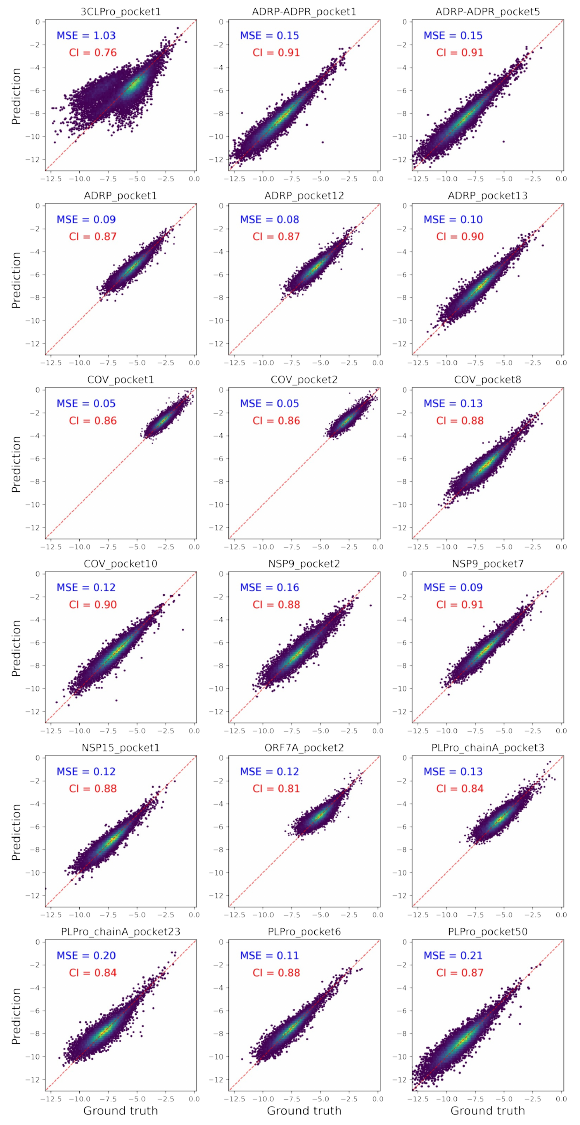}
    \caption{\label{fig_s:mse} Prediction performance on the test sets by the GatedGCN model. MSE stands for mean squared error and CI for concordance index.
    }
\end{figure}

\vspace{\baselineskip}

\begin{figure}[ht]
    \centering
    \includegraphics[width=0.9 \hsize]{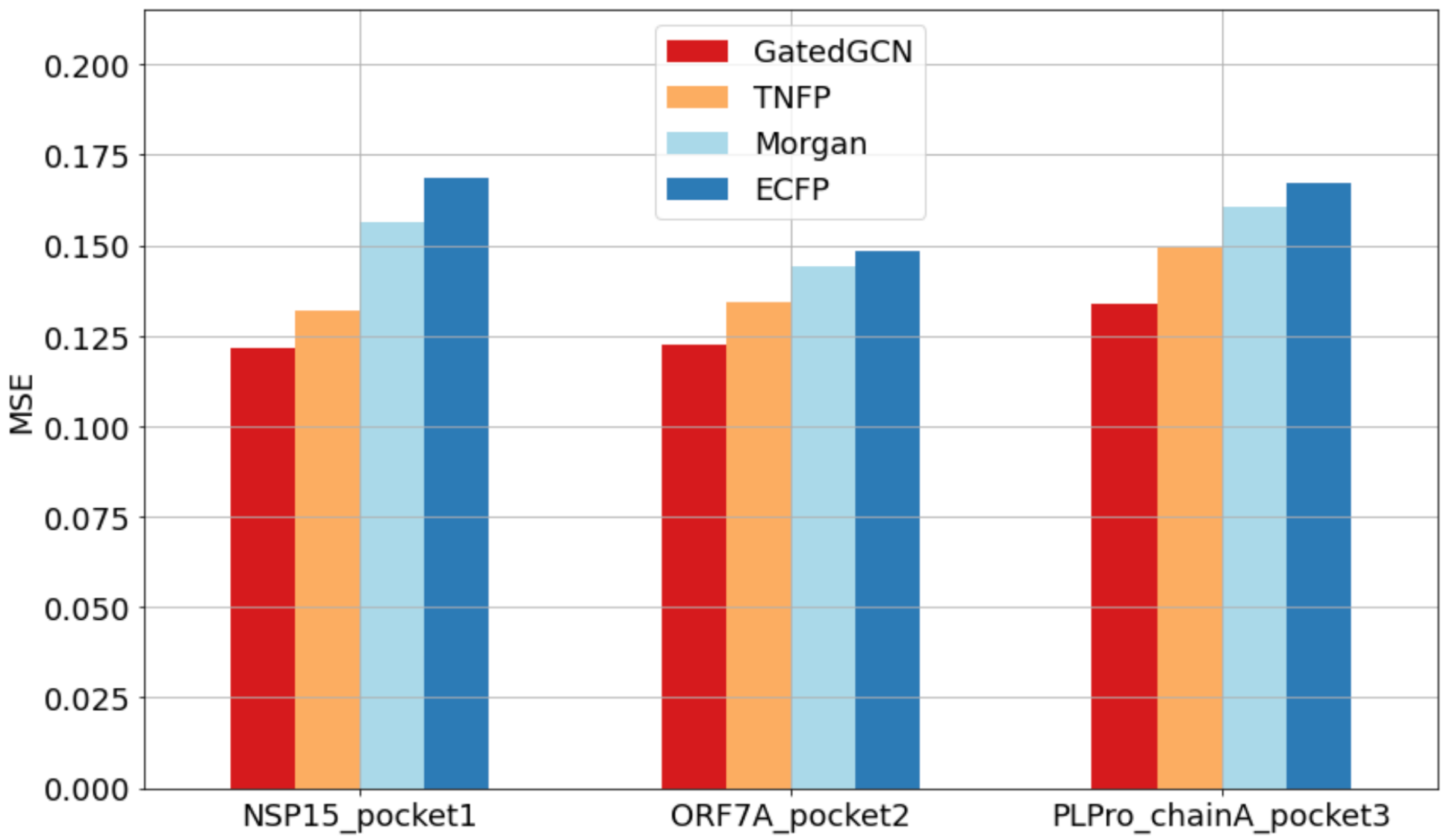}
    \caption{\label{fig:comp_perf} Performance comparison of the TNFP model with the GatedGCN model (Fig.~2b) and two CFP models (Fig.~2a) tested on three targets. 
    }
\end{figure}

\begin{figure}[htb]
    \centering
    \includegraphics[width=0.9 \hsize]{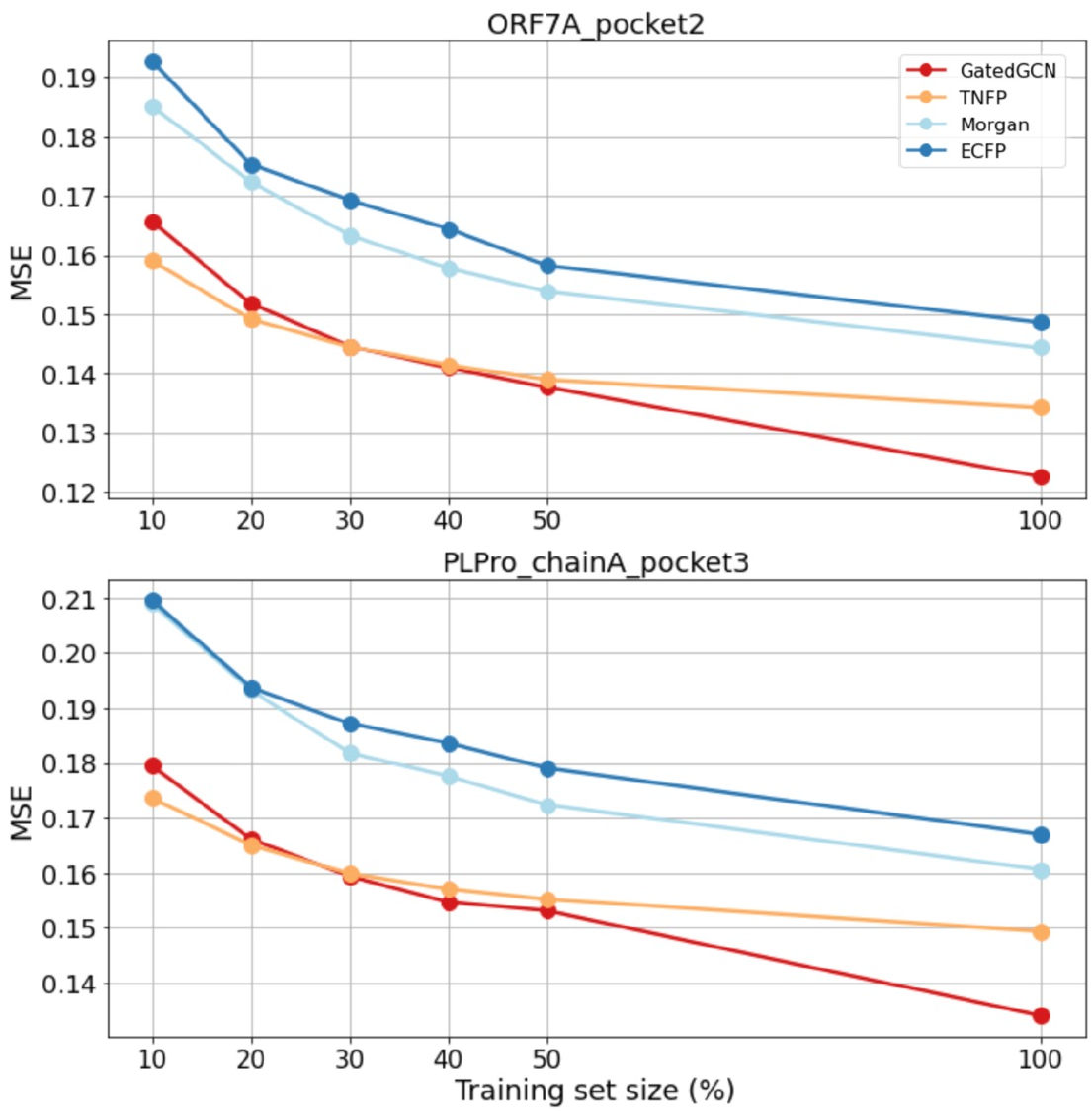}
    \caption{\label{fig_s:data_eff} Training data efficiency of four different models shown for two test targets.
    }
\end{figure}

\clearpage
%